\documentclass[pdflatex, sn-nature, iicol]{sn-jnl}

\usepackage[utf8]{inputenc}
\usepackage{graphicx,float}
\usepackage[makeroom]{cancel}
\usepackage[dvipsnames]{xcolor}
\usepackage{amsmath,amsfonts,amssymb,mathtools,amsthm}
\usepackage{caption}
\usepackage{algorithm,algorithmic}
\usepackage{subcaption}
\usepackage{hyperref}
\usepackage[most]{tcolorbox}
\usepackage{booktabs}
\usepackage{multirow}

\newcommand{\bzero}{\mathbf{0}}
  
\newcommand{\by}{\mathbf{y}}
\newcommand{\bx}{\mathbf{x}}
\newcommand{\bu}{\mathbf{u}}

\newcommand{\bd}{\mathbf{d}}
\newcommand{\bff}{\mathbf{f}}

\newcommand{\bg}{\mathbf{g}}
\newcommand{\bb}{\mathbf{b}}
\newcommand{\bq}{\mathbf{q}}
\newcommand{\br}{\mathbf{r}}

\newcommand{\bI}{\mathbf{I}}
\newcommand{\bA}{\mathbf{A}}
\newcommand{\bD}{\mathbf{D}}
\newcommand{\bC}{\mathbf{C}}
\newcommand{\bH}{\mathbf{H}}
\newcommand{\bQ}{\mathbf{Q}}
\newcommand{\bR}{\mathbf{R}}

\newcommand{\nin}{\noindent}

\newcommand{\btheta}{\boldsymbol{\theta}}
\newcommand{\bTheta}{\boldsymbol{\Theta}}
\newcommand{\bSigma}{\boldsymbol{\Sigma}}
\newcommand{\bphi}{\boldsymbol{\phi}}
\newcommand{\bmu}{\boldsymbol{\mu}}
\newcommand{\balpha}{\boldsymbol{\alpha}}
\newcommand{\bbeta}{\boldsymbol{\beta}}
\newcommand{\be}{\boldsymbol{e}}
\newcommand{\bE}{\mathbf{E}}

\newtheorem{approximation}{Approximation}

\begin{document}
\title{Truncated Neural Likelihood Estimation for Simulation-Based Inference in State-Space Models}

\author*[1]{\fnm{Kostas} \sur{Tsampourakis}}\email{kostas.tsampourakis@gmail.com}

\author[2]{\fnm{Víctor} \sur{Elvira}}\email{victor.elvira@ed.ac.uk}

\affil[1]{\orgdiv{School of Mathematics}, \orgname{University of Edinburgh}, \orgaddress{\street{}, \postcode{EH9 3FD}, \country{U.K.}}}

\affil[2]{\orgdiv{School of Mathematics}, \orgname{University of Edinburgh}, \orgaddress{\street{}, \postcode{EH9 3FD}, \country{U.K.}}}

\abstract{
State-space models (SSMs) are powerful probabilistic tools for modeling time-varying systems with latent dynamics. Inference in SSMs involves the estimation of latent states and parameters. In this work, we focus on parameter inference, which for SSMs is in general a very challenging problem due to the intractability of the likelihood. Recently, neural estimation methods, such as sequential neural likelihood (SNL), have shown promising results in Bayesian inference problems. In this paper, we show that SNL, when applied to the SSM setting, suffers important limitations, such as requiring a large amount of simulated samples to achieve a moderate performance, scaling poorly with sequence length, while not being amortized. We then introduce a novel inference algorithm called truncated-SNL (T-SNL), which addresses the limitations of SNL. Our algorithm is more accurate, more stable and robust during training, more scalable to longer temporal sequences, and can be amortized when new observations become available. Our experiments show that T-SNL is sample-efficient, robust, and flexible algorithm which outperforms other approaches. 
}


\maketitle

\section{Introduction}\label{sec1}
State-space models (SSMs) are probabilistic latent-variable models for time-series data \cite{sarkka2023bayesian}. A latent Markov process is used to model the state dynamics of the system. Information about the latent state is obtained through observations which are related to the state by a measurement model. The generality of SSMs makes them applicable to a wide range of problems across science and engineering, including robot state estimation and tracking \cite{ssm_robotics1, prob_robotics}, understanding and predicting states of neuronal ensembles in neuroscience \cite{paninski2010new, linderman2019hierarchical, aghagolzadeh2014latent}, tracking populations of interacting species in ecology \cite{buckland2004state, ssm_ecology2}, predicting social media evolution \cite{sharma2021recurrent}, analysing and predicting economic indices in econometrics \cite{hamilton1994state, lopes2011particle,sharma2021recurrent}, understanding mobility patterns \cite{martino2017cooperative}, and many others.

The two main tasks that arise when working with SSMs in the Bayesian setting, are the inference of states and parameters. State inference refers to the task of computing posterior distributions of latent states given a sequence of observations and the model parameters. This task is central in applications such as navigation, object tracking, and time series denoising. Parameter inference, on the other hand, deals with computing the posterior of the static parameters that define the dynamics and observation models, given the data. 

Both state and parameter inference in SSMs present computational challenges. The primary difficulty arises from the intractability of probability density integrals that arise from nonlinear transformations of non-Gaussian random variables. Approximate state inference methods such as Gaussian and particle filters have been developed to address this problem \cite{sarkka2023bayesian}. In this work, we focus on parameter inference which requires computation of an intractable likelihood. Many approximate parameter inference methods, such as EM \cite{shumway1982approach,chouzenoux2024sparse} and MCMC \cite{andrieu2010particle,kantas2015,luengo2020survey}, commonly use filters to obtain likelihood estimates recursively. Thus their accuracy is limited by the accuracy of the underlying filters. 

This is not the case for simulation-based inference (SBI) methods which avoid likelihood evaluations and instead rely on repeated simulations from the model. In contrast to standard methods, SBI works by simulating parameters and observations from the model, which are used to create a surrogate of the likelihood. This approach is particularly useful when the likelihood function is intractable or expensive to evaluate, but simulations from the model are easy to obtain. Independence from noisy likelihood estimates has made SBI a popular alternative to standard Bayesian inference tools \cite{sbi_atlas}.

The earliest SBI method is approximate Bayesian computation (ABC) \cite{tavare1997inferring}, which proposes parameter samples from the prior and accepts them if they generate simulated data sufficiently close to the observed data. Many extensions of ABC have been proposed since \cite{abchandbook, abcofthefuture}, most notably MCMC-ABC \cite{mcmc-abc} and SMC-ABC \cite{smc-abc, abc_tolerance}. More recently, methods based on neural density estimation \cite{made} have been proposed for the SBI problem, which work by using simulated samples to train neural network surrogates of the posterior, likelihood or likelihood ratio. These include methods such as sequential neural posterior (SNP) \cite{snpe-a, snpe-b, snpe-c}, sequential neural likelihood (SNL) \cite{snl}, and sequential neural ratio (SNR) \cite{sre-a, sre-b} which have become state-of-art in a variety of problems \cite{frontier, sbi_benchmark, sbi_atlas}. 

With the abundance of computation tools recently available to practitioners, such as GPUs and powerful software suites, SBI has emerged as an important subfield of Bayesian inference. It has been successfully applied to many real-world scenarios, with applications ranging from earthquake prediction \cite{earthquake_sbi}, economic agent-based models \cite{abc_econ1, abc_econ2}, astrophysics \cite{astro_sbi}, and biology \cite{sbi_biology}, to name a few. 

SBI methods have also been used successfully for parameter inference in SSMs \cite{toni2009approximate, abc_ssm, ssm_sbi, ssm_sbi2}. More recently there have been applications of neural SBI methods to SSM inference, in the context of VI and Gaussian processes \cite{neural_ssm_1, neural_ssm_2, neural_ssm_3}. While there have also been works using neural SBI for SSMs, these have mainly focused on inference on the hidden states of the SSM \cite{neural_ssm_4, neural_ssm_5}. Moreover, up to date and to the best of our knowledge, there have not been any works evaluating the performance of NDE-based methods such as SNL or SNPE, which are the best-performing methods, to SSM parameter inference. 

In this work, we introduce a novel method tailored to the parameter inference problem of SSMs. Our method addresses the drawbacks of SNL when applied to SSMs. We show that while SNL outperforms widely used algorithms such as SMC-ABC and MCMC, it suffers important limitations. For instance, it requires a large amount model simulations to achieve good performance, scales poorly with the length of the observation sequence, and is not amortized. We address these limitations by exploiting a well known forgetting property of SSMs \cite{kantas2015} to propose a novel, sample efficient algorithm, called truncated-SNL (T-SNL). T-SNL overcomes the limitations of SNL and achieves the best performance among the compared methods. The main contributions of this work are summarized as follows:

\begin{itemize}
    \item We propose truncated-SNL (T-SNL), a novel inference algorithm for SSMs. T-SNL addresses the limitations of SNL by leveraging the forgetting property of SSMs.
    \item We demonstrate that T-SNL is highly sample-efficient, stable, and amortized. We show that T-SNL consistently outperforms SNL, SMC-ABC, and particle MCMC across a range of state-space models.
\end{itemize}

The rest of the paper is organized as follows. In Section \ref{sec2}, we review the basic concepts used throughout the paper and introduce notation. In Section \ref{sec3}, we propose T-SNL and detail the methodology. In Section \ref{sec4}, we present the main experimental results of this work and discuss our findings. 

\section{Background}\label{sec2}
In this section, we introduce the main models and notations that set the stage for our contributions. 

\subsection{State-space models}\label{sec2.1}

State-space models (SSM) are probabilistic models that describe the time evolution of an indirectly observed system. The system at time $t$ is described by a latent \textit{state vector} $\bx_t$ in a \textit{state space} $\mathcal{X}$. The evolution of the state vector is given by a Markov chain in $\mathcal{X}$. The kernel of the Markov chain is known as the \textit{transition model} and is used to characterize the dynamical process.
Information about the system at time $t$ is obtained in the form of an \textit{observation vector} $\by_t$ in a space $\mathcal{Y}$. The conditional distribution of the observation vector given the state vector, known as the \textit{observation (or measurement) model}, is used to characterize the observation process.

Formally a SSM is described by 

\begin{align}
    \btheta &\sim p(\btheta), \\
    \bx_t &\sim p_{\btheta}(\bx_t | \bx_{t-1}),\ t=1,\dots,T, \\
    \by_t &\sim p_{\btheta}(\by_t | \bx_{t}),\ t=1,\dots,T,
\end{align}

\noindent where $p_{\btheta}(\bx_t | \bx_{t-1})$ is the transition model, $p_{\btheta}(\by_t | \bx_{t})$ the observation model, and we assume $\bx_0 \sim p_{\btheta}(\bx_0)$. The vector $\btheta \in \bTheta$ with prior $p(\btheta)$, encompasses the parameters of the transition and observation models of the SSM. 

Typically, to perform inference on the parameter vector $\btheta$, we need to evaluate the likelihood $p(\by_{1:T}|\btheta)$ for given data $\by_{1:T} = [\by_1,\dots,\by_T]$ and parameter $\btheta$. The likelihood of a SSM is given by

\begin{align}\label{eq:likelihood}
     p(\by_{1:T}|\btheta) = \int p(\bx_{0:T}, \by_{1:T}|\btheta) d\bx_{0:T},
\end{align}

\noindent where

\begin{align}\label{eq:joint}
    p(\bx_{0:T}, \by_{1:T}|\btheta) = p_{\btheta}(\bx_0)\prod_{t=1}^T p_{\btheta}(\bx_t | \bx_{t-1}) p_{\btheta}(\by_t | \bx_{t}),
\end{align}

\noindent is the joint density of latent states and observations given the parameters. 

The likelihood of a SSM, given in Eq. \eqref{eq:likelihood}, is an integral with respect to the latent states of the model. The integral can be computed exactly for linear-Gaussian SSMs by the Kalman filter. For nonlinear and non-Gaussian SSMs it is intractable, requiring numerical approximations. 
The most common approach for approximating the likelihood of a general nonlinear SSM is particle filtering \cite{sarkka2023bayesian, doucet2009tutorial}. 

It is possible to write the SSM likelihood in an alternative way as a product of all autoregressive conditionals as
\begin{equation}\label{auto-regressive factorization}
    p(\by_{1:T}|\btheta) = \prod_{t=1}^T p(\by_t|\by_{1:t-1}, \btheta),
\end{equation}
where each conditional $p(\by_t|\by_{1:t-1}, \btheta)$ is an intractable integral. In filtering, the factors are computed as normalization constants of the filtering distribution and the factorization of Eq. \eqref{auto-regressive factorization} is used to approximate the likelihood. 

\noindent\textbf{Forgetting properties of SSMs:} Many state-space models possess the \textit{exponential forgetting property} \cite{delmoral_feynman} which states that, for any $\bx_0, \bx'_0 \in \mathcal{X}$, there exist constants $B\in(0,\infty)$ and $\lambda\in[0,1)$ such that
\begin{equation}\label{eq:forg_prop}
    || p(\bx_t|\by_{1:t}, \bx_0) - p(\bx_t|\by_{1:t}, \bx'_0) ||_{TV} \leq B\lambda^t,
\end{equation}
where $p(\bx_t|\by_{1:t}, \bx_0)$ is the optimal filtering distribution at time $t$ when initialized at $\bx_0$, and $||\cdot||_{TV}$ is the total variation distance. This property is satisfied when the state process is uniformly ergodic and the observations satisfy certain conditions that can be found in \cite{kantas2015, delmoral_feynman}. Intuitively this means that the optimal filter forgets its initial condition exponentially fast. Equivalently it means that the influence of past observations on the filtering distribution is forgotten exponentially fast \cite{tsampourakis2022approximating}.

\subsection{Simulation-based inference}

Simulation-based inference (SBI) methods have been developed to perform inference on simulator-based models \cite{frontier}. Unlike traditional methods, which rely on a computable representation of the likelihood (e.g., in the form of an intractable integral), SBI methods perform multiple simulations from the model, which are used to obtain approximations of the likelihood. SBI methods simulate parameter-observation pairs $(\btheta^{(n)}, \by^{(n)})$ and use them together with the true observations $\by$ to learn a density estimator of the joint distribution $p(\btheta, \by)$, or the likelihood $p(\by|\btheta)$. 

For example, ABC methods propose parameter values repeatedly and use the simulator to generate synthetic data vectors \cite{abchandbook}. In the simplest version of ABC, known as \textit{rejection ABC}, a candidate parameter is sampled from the model prior and is used to simulate a synthetic dataset. The distance between the synthetic dataset and the observations is quantified using an appropriate metric (e.g. Euclidean distance). If the distance is larger than the tolerance, we reject the parameter value and repeat the procedure. We accept parameters that produce datasets \textit{close} to the observations. When the tolerance is set to $\epsilon$, the rejection ABC sampler works by repeating the following steps until the required sample sized has been reached:

\begin{itemize}
    \item[] \textbf{1.} sample candidate: $\btheta^{(n)} \sim p(\btheta)$,  \\
    \item[] \textbf{2.} simulate dataset: $\by^{(n)} \sim p(\by|\btheta)$, \\
    \item[] \textbf{3.} accept $\btheta$ if $d(\by^{(n)}, \by) \leq \epsilon$.
\end{itemize}

\nin In this way, the ABC rejection sampler produces samples from an approximation of the posterior. Decreasing the tolerance $\epsilon$ improves the approximation, in fact, ABC methods have been shown to asymptotically converge to the true posterior as the tolerance goes to zero \cite{frazier2018asymptotic}. In practice, however, a very small $\epsilon$ will lead to a vanishing acceptance rate, requiring an ever-increasing number of simulations to achieve a good posterior approximation. For rejection ABC, this problem is especially pronounced since the proposal mechanism is fixed. Since the acceptance rate depends on the proposal mechanism as well as $\epsilon$, improved algorithms adapt the proposal to achieve a non-vanishing acceptance rate while decreasing the tolerance. Notably, the ABC-MCMC \cite{mcmc-abc} algorithm uses the ABC sampler within an MCMC proposal kernel to perform likelihood-free MCMC. SMC-ABC \cite{smc-abc} uses SMC to target intermediate ABC-posteriors for a decreasing tolerance sequence while adapting the proposal to achieve a non-vanishing acceptance rate.

\subsection{Neural density estimation}

Neural density estimation (NDE) methods use neural network models to parametrize a data distribution \cite{made}. The model is trained on the observed dataset typically by maximizing the average log-likelihood using stochastic-gradient optimization. NDE methods have also been developed for conditional density estimation by defining a neural conditional density and training on pairs of datapoints \cite{bishop1994mixture}. Recently, NDE methods have been applied to SBI and have been shown to be accurate and flexible on this problem \cite{snpe-a, snl, frontier}. We briefly review the MADE \cite{made} and MAF \cite{maf} models that are used in this work. 

\subsubsection{MADE}

The \textit{masked autoencoder for distribution estimation} (MADE) model \cite{made} is an autoregressive density estimator. Given samples from a variable $\by$ in $\mathbb{R}^{D}$, we are interested in learning their density $p(\by)$. Autoregressive density estimators use the identity

\begin{equation}
    p(\by) = \prod_{d=1}^D p(y_d|y_{1:d-1})
\end{equation}

\nin and use neural networks to parametrize the conditionals. One common choice is to use univariate Gaussian conditionals, 

\begin{equation} \label{eq:gauss_conditional}
    p(y_d|y_{1:d-1}) = \mathcal{N}(y_d| \mu_d, \sigma_d^2),
\end{equation}

\nin and feedforward neural networks to model the mean and variance

\begin{align}
    \mu_d = g_{\mu_d}(y_{1:d-1};\bphi_{\mu_d}), \\
    \sigma_d = g_{\sigma_d}(y_{1:d-1};\bphi_{\sigma_d}),
\end{align}

\nin where $\bphi_{\mu_d}$ and $\bphi_{\sigma_d}$ are trainable parameters of the neural networks $g_{\mu_d}$ and $g_{\sigma_d}$, respectively, for $d=1,\dots,D$.

Instead of using $D$ separate neural networks to model the conditionals, MADE uses masking to compute all means and covariances using a single neural network:

\begin{align}
    (\mu_{1:D}, \sigma_{1:D}) = \bg_{\text{\tiny MADE}}(\by; \bphi),
\end{align}

\noindent where $\bphi$ are all trainable parameters. The function $\bg_{\text{\tiny MADE}}$ is a MLP, where the weight matrix of each layer is multiplied by a binary mask which drops some connections, as introduced in \cite{made}. This ensures that $\mu_d$ and $\sigma_d$ only depend on $y_{1:d-1}$. 

\subsubsection{MAF}

Autoregressive models such as MADE can also be viewed as normalizing flows. We can see this by writing Eq. 

\eqref{eq:gauss_conditional} as
\begin{equation}\label{eq:made_flow_transformation}
    y_d = u_d\sigma_d(y_{1:d-1}) + \mu_d(y_{1:d-1}),
\end{equation}

\noindent with $u_d \sim \mathcal{N}(0,1)$, for $d=1,\dots,D$. This defines an invertible transformation $\by = \bff(\bu)$, where $\bu \sim \mathcal{N}(\bzero, \bI)$. The inverse of this transformation $\bu = \bff^{-1}(\by)$ is given elementwise by

\begin{equation}\label{eq:made_flow_inverse}
    u_d = (y_d - \mu_d(y_{1:d-1}))\sigma_d(y_{1:d-1})^{-1}.
\end{equation}

The Jacobian of this transformation is triangular due to the autoregressive property, and its determinant is given by

\begin{equation}
    \Big|\det\frac{\partial \bff^{-1}}{\partial\by}\Big| = \prod_{d=1}^D \frac{1}{\sigma_d}.
\end{equation}

\nin The MAF is constructed by stacking multiple MADE flows \cite{maf}. Stacking $K$ MADE flows we obtain the 

\begin{equation}
    \by = \bff_{K}\circ\cdots\bff_1(\bu),
\end{equation}

\nin where each $\bff_k$ is a MADE flow transformation as the one in Eq. \eqref{eq:made_flow_transformation}, with inverse as the one in Eq. \eqref{eq:made_flow_inverse}. This results in a very flexible model that can be trained efficiently by standard gradient methods. Moreover, $p(\by)$ can be calculated by

\begin{equation}
    p(\by) = p(\bu) \Big|\det\frac{\partial\bff_{\text{ \tiny MAF}}^{-1}}{\partial\by}\Big|,
\end{equation}

\nin where $\bff_{\text{ \tiny MAF}} = \bff_{K}\circ\cdots\bff_1$, and 

\begin{equation}
    \Big|\det\frac{\partial\bff_{\text{ \tiny MAF}}^{-1}}{\partial\by}\Big| = \prod_{k=1}^K\prod_{d=1}^D \frac{1}{\sigma^{(k)}_d},
\end{equation}

\nin where $\sigma_{1:D}^{(k)}$ are scale parameters computed by the $k^{th}$ MADE layer.

\noindent\textbf{Conditional NDE. }Autoregressive models such as MADE and MAF naturally extend to conditional density estimation, i.e., the problem of estimating the conditional density $p(\by|\btheta)$. This can be done using NDE, by treating $\btheta$ as the leading dimensions of an augmented data vector $(\btheta, \by)$ and only modeling the conditionals that correspond to $\by$ \cite{maf}. 

\subsubsection{Using neural density estimation for SBI}

Neural density estimation methods naturally lend themselves to the problem of SBI, since they can be employed to model unknown probability densities. There are two main approaches using NDE for this problem, targeting the posterior or likelihood of the model. \textit{Neural posterior estimation} methods use NDE models to learn a neural estimate of the posterior \cite{snpe-a, snpe-b} by training on proposed parameter-data pairs. \textit{Neural likelihood estimation} methods learn a neural estimate of the likelihood \cite{snl} by training on generated samples, targeting the conditional $p(\by|\btheta)$. 

\begin{figure}
     \centering
     \begin{subfigure}[b]{\columnwidth}
         \centering \includegraphics[width=\columnwidth]{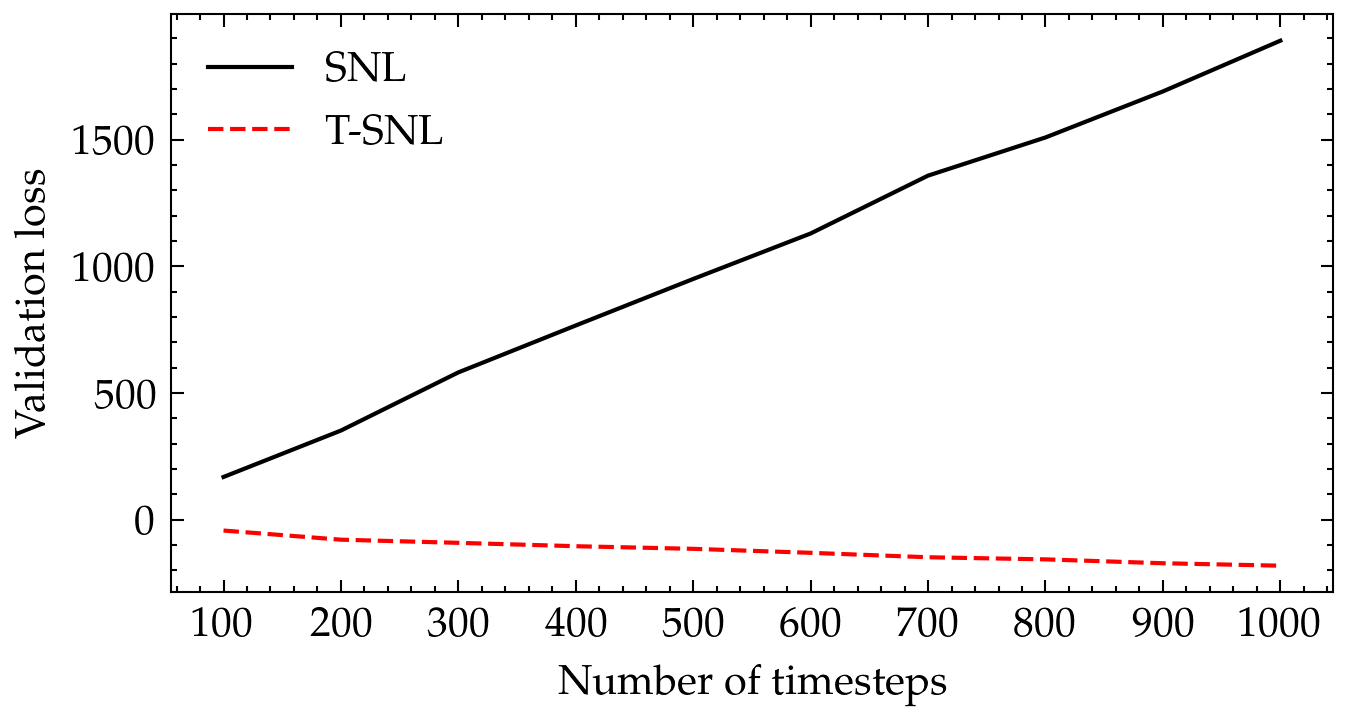}
     \end{subfigure}
     \hfill
    \caption{Validation losses of SNL and T-SNL during training for an increasing number of simulation timesteps. While for SNL the loss increases with the number of timesteps, for T-SNL it remains constant.}
    \label{fig:val_loss}
\end{figure} 

\section{Neural likelihood estimation for inference in SSMs}\label{sec3}

In this section, we introduce a novel, sample-efficient algorithm which is tailored to the SSM inference problem. Our algorithm, denoted truncated-SNL (T-SNL) learns a neural surrogate of the model likelihood sequentially. T-SNL works by replacing the likelihood factors $p(\by_t|\by_{1:t-1})$ of Eq. \eqref{auto-regressive factorization} by a truncated version $p(\by_t|\by_{t-L:t-1})$ which only conditions on the $L$ most recent observations. This is equivalent to the assumption that the observation process is Markovian of order $L$. Intuitively, this is convenient because it reduces the problem of learning $T$ different conditionals $\{p(\by_t|\by_{1:t-1})\}_{t=1}^T$ to that of learning a single conditional, namely $p(\by_t|\by_{t-L:t-1})$. This trick allows us to obtain a dataset for training T-SNL that is $T$ times larger than the SNL dataset, for the same number of simulations. This makes the T-SNL algorithm significantly more sample efficient than SNL, as we show numerically in the next section. 

\subsection{Truncation of likelihood factors}

This section proceeds by motivating the truncation of the likelihood factors, before introducing the T-SNL algorithm in detail. A key limitation of SNL when applied to state space models (SSMs) is that it approximates the full data likelihood $p(\by_{1:T}|\btheta)$ directly via simulations. Due to the autoregressive structure of the SSM likelihood, this effectively amounts to learning $T$ distinct conditional distributions of the form $p(\by_t|\by_{1:t-1},\btheta)$, one for each time step. While this strategy is widespread, it is inefficient: the conditionals share structure, yet SNL does not reuse information across time steps. As a result, each simulation yields only one training point for the surrogate model, limiting sample efficiency.

This redundancy becomes especially apparent in light of the exponential forgetting property of SSMs. As discussed in Sec. \ref{sec2.1}, Eq. \eqref{eq:forg_prop}, this property implies that if we initialize two filters with different initial conditions and they process the the same data, they will quickly converge to the same distribution. This implies that the sequence of likelihood factors, which is expressed in terms of the filtering distribution as 

\begin{align*}
p(\by_t|\by_{1:t-1}
) =\int p_{\btheta}(\by_t|\bx_t) p_{\btheta}(\bx_t|\bx_{t-1}) \times\\
\times p(\bx_{t-1}|\by_{1:t-1})d\bx_{t-1:t},
\end{align*}

\noindent depends strongly on more recent observations and is independent of observations in the distant past.

This observation motivates a more efficient approach: rather than learning $T$ separate conditional distributions, we approximate each likelihood factor by a truncated version that conditions only on the $L$ most recent observations. Specifically, we assume that

\begin{equation}
p(\by_t|\by_{1:t-1}, \btheta) \approx p(\by_t|\by_{t-L:t-1}, \btheta),
\end{equation}

\nin which is equivalent to approximating the observation process as an $L$-order Markov chain. Under this approximation, all the conditional factors in the likelihood share the same structure and can be modeled using a single neural density estimator. This is formalized in the following approximation.

\begin{approximation}[Observation process is approximately Markovian]\label{approximation 1}
We approximate the observation process by a Markov chain of order $L$. The kernel of the chain has conditional density $p(\by_t|\by_{t-L:t-1},\btheta)$, such that the likelihood can be approximated as
\begin{equation}\label{eq:truncated_likelihood_factorization}
p_L(\by_{1:T}|\btheta) \approx \prod_{t=1}^T p(\by_t|\by_{t-L:t-1},\btheta).
\end{equation}
\end{approximation}

This truncated factorization greatly improves sample efficiency. Given a dataset $\mathcal{D} = \{(\btheta^{(n)}, \by^{(n)}_{1:T})\}_{n=1}^N$ consisting of $N$ simulations, the kernel is obtained by minimizing the log-likelihood on the lagged set $\mathcal{D}_L=\{\{(\btheta^{(n)}, \by^{(n)}_{t-L:t})\}_{t=1}^T\}_{n=1}^N$. By comparison SNL, which targets the full likelihood, is trained on $\mathcal{D}$. Thus SNL uses a dataset of size $|\mathcal{D}|=N$, while T-SNL a set of size $|\mathcal{D}_L|=N \times T$. This multiplication of the dataset size is a main strength of T-SNL which has the remarkable property that for some scenarios, it can tackle the inference problem with a single simulator run, $N=1$. This makes T-SNL a very sample-efficient algorithm, as we will show experimentally in Section \ref{sec4}. We continue with a detailed description of the proposed methodology.

\subsection{The T-SNL algorithm}

We now present the full truncated-SNL (T-SNL) algorithm for parameter inference in state-space models. T-SNL replaces the full likelihood with a truncated factorization that is easier to estimate and more sample-efficient. The learned likelihood is used within an MCMC sampler to approximate the posterior. T-SNL uses sequential proposal adaptation, which allows the algorithm to focus simulations on high-probability regions of the parameter space.

\subsubsection{Description of the algorithm}

The T-SNL algorithm is detailed in Alg. \ref{alg:T-SNL}. First, the parameter proposal is initialized as the model prior. Then, for each round $r = 0, \ldots, R-1$, parameter samples are drawn from the current proposal $\pi_r$ via MCMC in Eq. \eqref{eq:T-SNL_prop_param}. Observation sequences are drawn from the simulator as shown in Eq. \eqref{eq:T-SNL_prop_obs}. 

From the simulations a dataset $\mathcal{D}_L$ is constructed in Eq. \eqref{eq:lagged_ds}. The training dataset $\mathcal{D}_{r+1}$ for training in round $r+1$ is then built according to one of three strategies $\textsc{all}$, $\textsc{last}$, or $\textsc{best}$. $\textsc{all}$ means that all past simulations are used, $\textsc{last}$ that only the simulations from the current round are used, and $\textsc{best}$ means that the $N$ datapoints from all rounds whose trajectories are closest to the observed data (in Euclidean distance)

Then, the likelihood model is trained. A MAF model $q^{(L)}_{\bphi}(\by_t|\by_{t-L:t-1}, \btheta)$ is used to model the Markov kernel of the observation process. The model is trained by maximizing the log-likelihood over the dataset $\mathcal{D}_{r+1}$. More specifically, we obtain $\bphi$ by minimizing the loss function,
\begin{align}
   \text{loss}(\bphi) &= -\sum_n \log q^{(L)}_{\bphi}(\by_{1:T}^{(n)}|\btheta^{(n)}) \\ 
    &= -\sum_{n=1}^N \sum_{t=1}^T \log q^{(L)}_{\bphi}(\by_t^{(n)}|\by_{t-L:t-1}^{(n)}, \btheta^{(n)}).
\end{align}

\noindent The full likelihood estimate is given in Eq. \eqref{eq:T-SNL_likelihood}.

The learned likelihood is used to form the approximate posterior $\widehat{p}(\btheta|\by_{1:T}^{(obs)})$, given in Eq. \eqref{eq:T-SNL_approx_posterior}. This approximation is used as the proposal $\pi_{r+1}(\btheta)$ in the next round.

\subsection{Choice of lag $L$}

To select a value for the lag $L$ of the T-SNL algorithm we estimate the autocorrelation function (ACF) of the observation sequence. We choose $L$ that attains a small ACF. Approximation \ref{approximation 1} implies that $\by_{t-k}, \by_t$ should be independent when $k>L$, hence the ACF should be low. 


\begin{center}
\scalebox{1.0}{
\begin{minipage}{\linewidth}
\begin{algorithm}[H]
\begin{algorithmic}[1]
\STATE \textbf{Initialization} Set $\pi_0(\btheta) = p(\btheta)$ and $\mathcal{D}_0=\{\}$
\STATE \textbf{For $r=0,\dots,R-1$:} \\
\STATE \textbf{Simulation:} Sample parameters from the proposal and observations from the SSM:
\begin{align} \label{eq:T-SNL_prop_param}
    \btheta^{(n)} &\sim \pi_{r}(\btheta), \\ \label{eq:T-SNL_prop_obs}
    \by_{1:T}^{(n)} &\sim p(\by_{1:T}|\btheta^{(n)}),
\end{align}
for $n=1,\dots,N$. 
Set up the lagged dataset 
\begin{equation}\label{eq:lagged_ds}
    \mathcal{D}_L=\{\{(\btheta^{(n)}, \by^{(n)}_{t-L:t})\}_{t=1}^T\}_{n=1}^N.
\end{equation}
\STATE \textbf{Set up dataset:} 
One of three options is used to construct the training dataset at round $r$:
\begin{itemize}
    \item[-] \textsc{all}:  $\mathcal{D}_{r+1} = \mathcal{D}_{r} \cup \mathcal{D}_L$
    \item[-] \textsc{last}:  $\mathcal{D}_{r+1} = \mathcal{D}_L$
    \item[-] \textsc{best}: $\mathcal{D}_{r+1}$ is the set of the $N$ datapoints from all rounds closest to the observations.
\end{itemize}

\STATE \textbf{Training:} Train $q^{(L)}_{\bphi}(\by_t|\by_{t-L:t-1}, \btheta)$ on $\mathcal{D}_{r+1}$ and compute likelihood approximation \\
\begin{equation}\label{eq:T-SNL_likelihood}
    q^{(L)}_{\bphi}(\by_{1:T}^{(obs)}|\btheta) = \prod_{t=1}^T q^{(L)}_{\bphi}(\by_t^{(obs)}|\by_{t-L:t-1}^{(obs)}, \btheta).
\end{equation}
\STATE \textbf{Posterior approximation:} Set\\
\begin{align}\label{eq:T-SNL_approx_posterior}
    \widehat p(\btheta|\by_{1:T}^{(obs)}) & \propto q^{(L)}_{\bphi}(\by_{1:T}^{(obs)}|\btheta) p(\btheta), \\
    \pi_{r+1}(\btheta) & = \widehat p(\btheta|\by_{1:T}^{(obs)}).
\end{align}
\end{algorithmic}

\caption{Truncated SNL (T-SNL)}
\label{alg:T-SNL}
\end{algorithm}
\end{minipage}%
}
\end{center}


\vspace{1cm}

In particular, for a sequence of observations $(\by_t)_{t=1}^T$ and lag $L$, we calculate the autocorrelation matrix 

\begin{equation}
    \bC_L = \frac{1}{T-L} \sum_{t=1}^{T-L}(\by_{t-L} - \bar \by) (\by_{t} - \bar \by)^T ,
\end{equation}

\nin which is the sample autocorrelation matrix at lag $L$. The frobenius norm of this matrix,

\begin{equation}\label{acf}
    ||\bC_L||_F = \text{trace}(\bC_L^T\bC_L)^{1/2},
\end{equation}

\nin is an estimate for the total autocorrelation among all dimensions of $\by$. The ACF can be used as a rule of thumb for the determination of $L$.

\section{Numerical Experiments}\label{sec4}

We conduct a series of experiments comparing T-SNL against established inference methods on a range of state-space models of increasing complexity. Our goal is to assess the sample efficiency and robustness of the proposed algorithm and highlight its advantages over SNL in a variety of inference settings. We consider the LGSSM, stochastic volatility models with varying dimension, and an example with nonlinear population dynamics. For each setting, we measure the quality of posterior inference under a fixed simulation budget and compare the trade-off between accuracy and simulation cost across methods. In Sec. \ref{sec4.1} we describe the experimental setup, including the algorithms and evaluation metrics. The results are presented in Sec. \ref{sec4.2}.

\begin{figure}
     \centering
     \begin{subfigure}[b]{\columnwidth}
         \centering \includegraphics[width=\columnwidth]{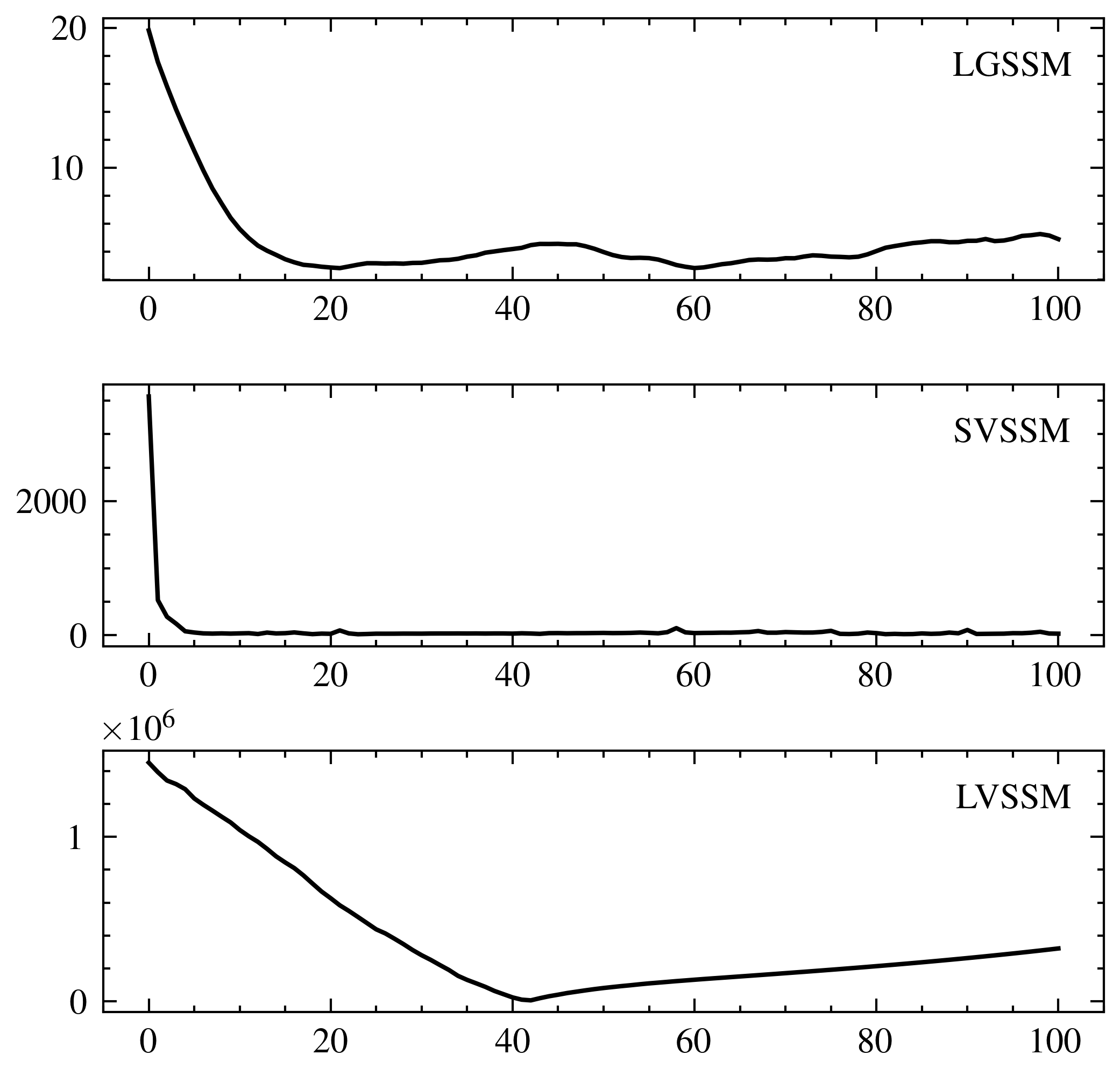}
     \end{subfigure}
     \caption{\textbf{ACF.} Autocorrelation function for each of the state-space models used in this work. Each plot is obtained by averaging multiple ACF estimates.}
     \hfill
    \label{fig:acf}
\end{figure}

\subsection{Setup} \label{sec4.1}

\subsubsection{Algorithms.} 

In our experiments we evaluate and compare the following algorithms. 

\noindent \textbf{SMC-ABC.} We use SMC-ABC with a Gaussian kernel and resampling when the effective sample size is less than $50\%$. We use the adaptive method of \cite{abc_tolerance} to select the sequence of tolerances $\epsilon_n$ and the stopping rule. Details of our implementation of this method can be found in the supplementary material. 

\noindent \textbf{Particle MCMC.} Particle MCMC is a class of algorithms that embeds a particle filter within a Markov chain Monte Carlo kernel to allow sampling from the posterior when the likelihood is intractable but can be unbiasedly estimated \cite{andrieu2010particle}. In our implementation, a bootstrap particle filter (BPF) \cite{gordon1993novel} estimator of the model likelihood is used within a random walk Metropolis (RWM) kernel. We run the MCMC chain for 1000 steps.

\noindent \textbf{SNL \& T-SNL.}
For both SNL and T-SNL we use a MAF model composed of 5 MADE layers, each composed of 5 hidden layers, each of them with 32 hidden units and relu or tanh activations. After conducting several tests, we found those architectures to achieve the best model size to performance tradeoff. The models are trained using the adam optimizer \cite{kingma2014adam}. After the models are trained, we use RWM or elliptical slice sampling \cite{murray2010elliptical} to obtain samples from the posterior.

\subsubsection{Metrics.} 
We use the following commonly used metrics to measure the performance of the algorithms \cite{sbi_benchmark}. 

\noindent\textbf{Probability of true parameters} One metric that we used to assess the accuracy is the negative log probability of the true parameters under a kernel density estimate on the posterior samples. More precisely, if $\widehat \btheta_1, \dots, \widehat \btheta_K$ are samples from an approximate posterior, and $\btheta_0$ is the true parameter value we define the error as
\begin{equation}
    \mathcal{E}_{\text{\tiny KDE}} = - \log p_{\text{\tiny KDE}}(\btheta_0 | \widehat \btheta_1, \dots, \widehat \btheta_K),
\end{equation}
where $p_{\text{\tiny KDE}}(\btheta | \widehat \btheta_1, \dots, \widehat \btheta_K)$ is the pdf of a kernel density estimator with kernels placed at the sample points $\widehat \btheta_1, \dots, \widehat \btheta_K$. We use standard normal kernels for the evaluation of $\mathcal{E}_{\text{\tiny KDE}}$.

\noindent \textbf{Minimum distance.} Another informative measure of discrepancy  is the minimum distance between posterior samples and the true parameter
\begin{equation}
    \mathcal{E}_{\text{\tiny min}} = \min_k||\btheta_0 - \widehat \btheta_k||.
\end{equation}
In the supplementary material we show that $\mathcal{E}_{\text{\tiny KDE}}$ is  upper and lower bounded by functions of $\mathcal{E}_{\text{\tiny min}}$. Advantages of $\mathcal{E}_{\text{\tiny min}}$ over $\mathcal{E}_{\text{\tiny KDE}}$ are that it is independent of the choice of kernel density and that it has an intuitive meaning. 

\noindent \textbf{Posterior bias and standard deviation.} We also report the bias and standard deviation of the posterior samples:

\begin{align}
    \text{bias} &= ||\btheta_0 - \widehat \btheta||, \\
    \text{st.dev.} &=\sqrt{\frac{1}{K}\sum_{k=1}^K||\widehat \btheta_k - \widehat \btheta||^2},
\end{align}
    
\noindent where $\widehat \btheta = \sum_{k=1}^K\widehat \btheta_k / K$ is the sample mean. These quantities give us insight on the distribution of the samples. The bias informs about the distance between the sample mean and the true parameter, while the standard deviation about the spread of the samples around their mean. Together they give us the RMSE of the estimator which is equal to the sum of squares of bias and standard deviation. 


\noindent \textbf{Simulation cost.} To measure the amount of data that each algorithm uses to perform inference, i.e., the simulation cost for each algorithm, we count the number of calls to the dynamics simulator $p(\bx_t|\bx_{t-1})$. This is equal to the number of time steps times the number of simulations of the full model that each algorithm uses. We choose this measure because it can be applied to all considered algorithms and is a constant multiple of the number of samples, which is the usual metric.

\subsection{Results} \label{sec4.2}

\begin{figure}
     \centering
     \begin{subfigure}[b]{\columnwidth}
         \centering \includegraphics[width=\columnwidth]    {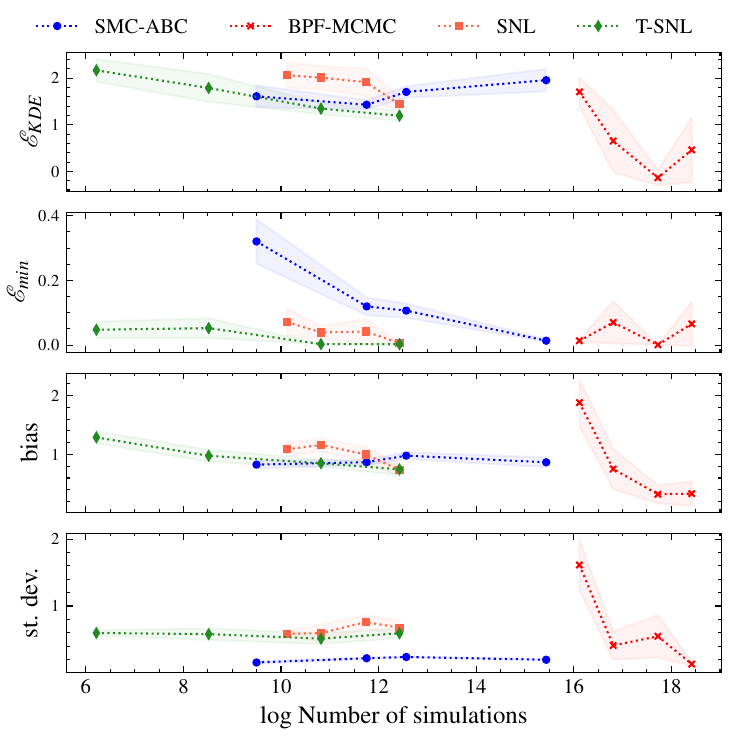}
     \end{subfigure}
     \hfill
    \caption{\textbf{LGSSM.} Results for the inference problem for the dynamics covariance. We plot errors vs number of simulations, bottom left is best.}
    \label{fig:lgssm1}
\end{figure}

\begin{figure}
     \centering
     \begin{subfigure}[b]{\columnwidth}
         \centering \includegraphics[width=\columnwidth]{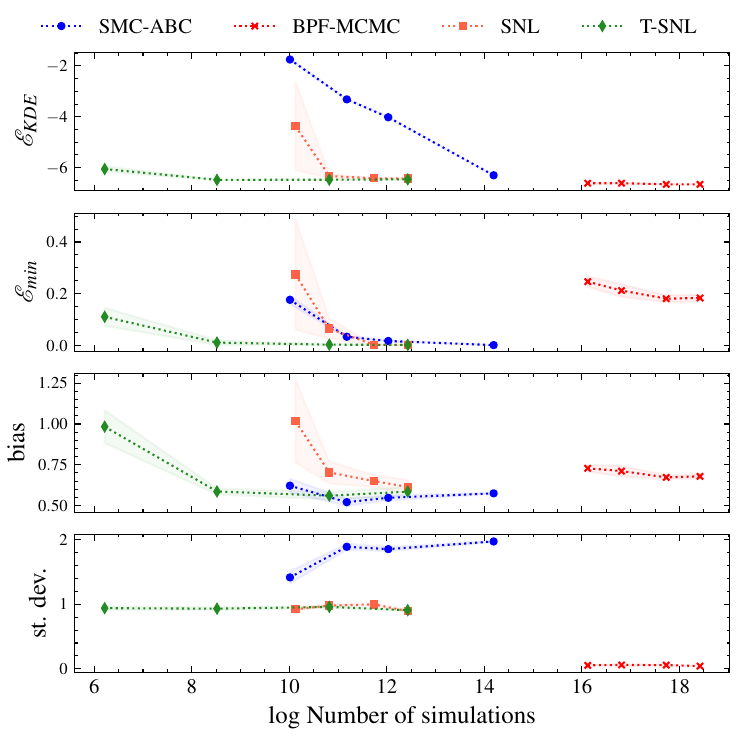}
     \end{subfigure}
     \hfill
    \caption{\textbf{SVSSM.} Results for the inference problem for the correlation matrix. We plot errors vs number of simulations, bottom left is best.}
    \label{fig:svssm1}
\end{figure}

\begin{figure}
     \centering
     \begin{subfigure}[b]{\columnwidth}
         \centering \includegraphics[width=\columnwidth]{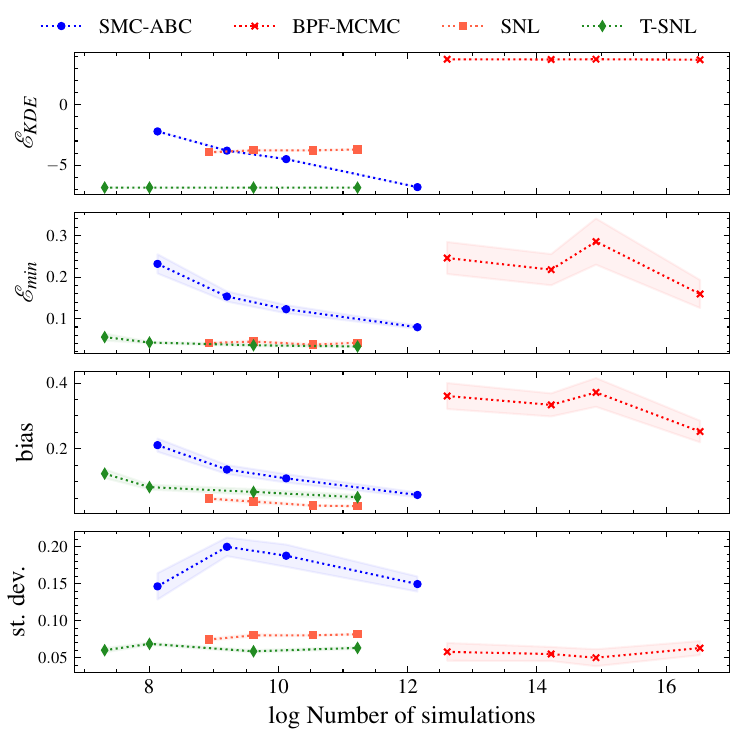}
     \end{subfigure}
     \hfill
    \caption{\textbf{LVSSM.} Results for the inference problem for the reaction rates. We plot errors vs number of simulations, bottom left is best.}
    \label{fig:lvssm1}
\end{figure}

We conduct experiments on the state-space models described below. For each SSM we consider scenarios in which we target different parameters during inference. In each experiment, we set the ground truth of the target parameters to a fixed value, then simulate observations and run inference for each algorithm, repeating for multiple trials. In Figs. \ref{fig:lgssm1}-\ref{fig:lvssm1} we plot the errors vs simulation cost averaged over trials.

\noindent \textbf{Linear Gaussian model.} The linear-Gaussian SSM is used to describe systems that evolve over time with linear dynamics and observation models and Gaussian noise. It is given by
\begin{align}
    \bx_t &= \bA \bx_{t-1} + \bq_t, \\
    \by_t &= \bH \bx_t + \br_t,
\end{align}
where $\bx_t\in\mathbb{R}^{d_x}$ and $\by_t\in\mathbb{R}^{d_y}$ are the state and observation vectors respectively, $\bA \in\mathbb{R}^{d_x\times d_x}$ and $\bH\in\mathbb{R}^{d_y\times d_x}$ are real matrices and $\bq_t \sim \mathcal{N}(\bq_0, \bQ)\in\mathbb{R}^{d_x}$ and $\br_t \sim \mathcal{N}(\br_0, \bR)\in\mathbb{R}^{d_y}$ are noise vectors. The initial state has distribution $\bx_0\sim\mathcal{N}(\bmu_0, \bSigma_0)$. 

\begin{figure}
     \centering
     \begin{subfigure}[b]{\columnwidth}
         \centering \includegraphics[width=\columnwidth]{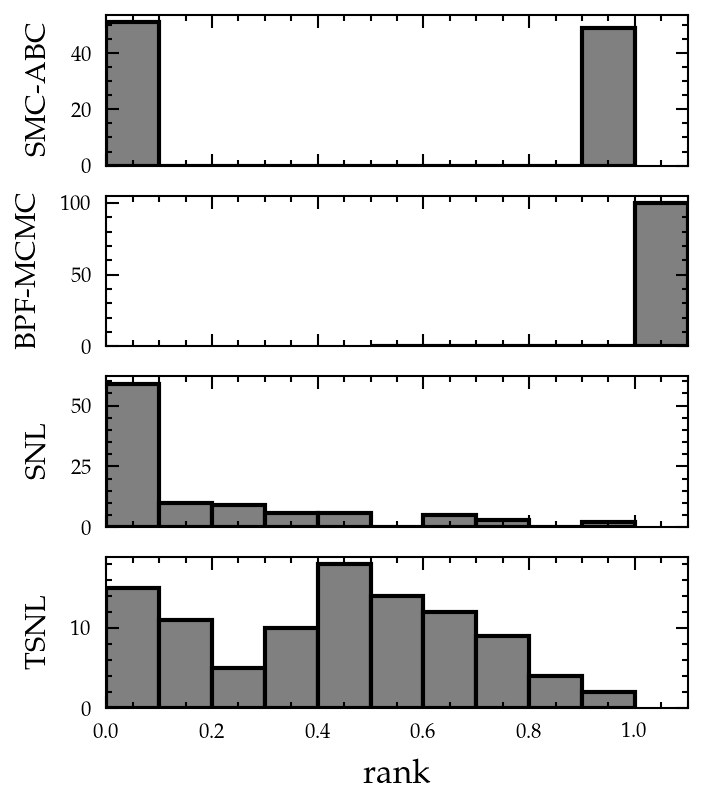}
     \end{subfigure}
     \hfill
     \caption{Rank statistics histograms for each of the algorithms tested in the LGSSM of Fig. \ref{fig:lgssm1}. A histogram that is closer to the uniform distribution indicates better calibration of the posterior. Each histogram plots 100 samples in 10 bins.}
    \label{fig:rank}
\end{figure}

We consider inference of the dynamics covariance $\bQ$ of a model with $d_x=d_y=1$, setting the ground truth to $\bQ_{gt} = 0.1 \bI$, with all other parameters known. The plots of metrics versus simulation cost are shown in Fig. \ref{fig:lgssm1}. We see that T-SNL achieves good performance using significantly fewer simulations than other methods. From the plot of $\mathcal{E}_{\text{\tiny KDE}}$, we see that BPF-MCMC is able to achieve the best accuracy, albeit at a much higher cost. Additionally, T-SNL provides better calibrated posteriors, as evidenced by the rank statistics of the true parameter versus posterior samples shown in Fig. \ref{fig:rank}. The histograms for T-SNL while not perfectly uniform are more even, while those of SMC-ABC, BPF-MCMC, and SNL show stronger skewness.

\noindent \textbf{Stochastic volatility model.} This model, which is prominent in econometrics, describes the time-evolution of coupled financial assets. The model is defined by

\begin{align}
    \bx_t &= \bA \bx_{t-1} + \bb + \bq_t, \\
    \by_t &= \bd + \bSigma_t^{1/2}\br_t,
\end{align}

\noindent where $\by_t$ is the vector of observed asset prices at time $t$ with covariance $\bSigma_t =\bD_t \bC \bD_t$ and $\bD_t$ is the matrix of volatilities which depends on the hidden state $\bx_t$ by $\bD_t = \text{diag}(e^{\bx_t / 2})$. Finally, $\bC$ is the matrix of correlations of $\by_t$, the vectors $\bb$ and $\bd$ are biases, and $\bq_t, \br_t$, are independent Gaussian noise vectors with covariances $\bQ$ and $\bR$ respectively. In Fig. \ref{fig:svssm1}, we present the results from the inference of matrix $\bC$ for a 2D model. 

We take the ground truth to be

\begin{equation}
    \bC_{gt} = \begin{pmatrix}
    1.00 & 0.52 \\
    0.52 & 1.00
    \end{pmatrix}.
\end{equation}

\noindent T-SNL exhibits the best error-to-cost tradeoff among the evaluated methods. It achieves similar or better accuracy than SNL while requiring significantly fewer simulations. In contrast, SNL performs well only after a larger number of simulations and tends to be noisy when data is limited. SMC-ABC shows high KDE error at low simulation budgets, which gradually improves as more simulations are added. The bias and standard deviation plots reveal distinct behaviors: BPF-MCMC tends to produce samples with low variance but persistent bias, suggesting that its estimates are tightly clustered but systematically miss the true parameter. In contrast, SMC-ABC produces samples with higher variance and smaller bias, which increases the chance of covering the true parameter. However, its variance remains considerably higher than that of both SNL and T-SNL, whose posteriors are more tightly concentrated around the ground truth.

\noindent \textbf{Lotka-Volterra model.}

The Lotka-Volterra model \cite{lotka_volterra} is a population model for two interacting populations of predator and prey. The stochastic version of the model can be expressed as a chemical reaction network with four reactions 

\begin{align*}
    A + B &\overset{\rho_1}{\longrightarrow} 2A, \\
    A + B &\overset{\rho_2}{\longrightarrow} A, 
    \\
    B &\overset{\rho_3}{\longrightarrow} 2B, \\
    A &\overset{\rho_4}{\longrightarrow} \emptyset,
\end{align*}

\nin where $A$ and $B$ are the predator and prey species respectively and $\rho_r>0$ is the rate of reaction $r=1,\dots,4$. The population dynamics of stochastic reaction networks can be simulated with the Gillespie algorithm \cite{gillespie1977exact}. Details of the model and the algorithm can be found in \cite{wilkinson2018stochastic}. 

The LV simulator begins at initial populations $(n_{A,0}, n_{B,0}) = (50, 100)$ and simulates a trajectory $\bx_t = (n_{A,t}, n_{B,t})$, where $n_{s,t}$ is the population of species $s\in\{A, B\}$ at $t=1,\dots,T$. We take the observations to be noisy measurements of the prey population

\begin{equation}
    y_t = n_{B,t} + e_{B,t},
\end{equation}

\nin where $e_{B,t}\sim\mathcal{N}(0,\sigma_e^2 \bI)$. 

Results are shown in Fig. \ref{fig:lvssm1}. The overall conclusion is similar to that of the other two models. T-SNL is again able to achieve the best tradeoff between error and simulation cost. SMC-ABC is able to achieve similar performance as the simulation cost increases. In contrast SNL does not seem able to improve the estimate for larger samples. BPF-MCMC produces again highly confident biased estimates. 


\subsubsection*{Discussion}

Our experiments show that T-SNL is an accurate and robust algorithm for parameter inference in state-space models, while requiring a very small number of training samples from the simulator. The comparison between T-SNL and SMC-ABC shows that while SMC-ABC can perform well on simple, low-dimensional models such as the LGSSM, it becomes very inefficient as the complexity of the simulator increases. In contrast the sample efficiency of T-SNL persists as simulator complexity increases, requiring a minimal number of samples to achieve near-optimal performance.

This is most evident in models with nonlinear dynamics, such as the stochastic volatility and Lotka-Volterra systems. In these cases, the rejection-based nature of SMC-ABC becomes increasingly inefficient, and tuning the tolerance schedule becomes critical. T-SNL, by contrast, scales naturally: its neural surrogate can handle complex likelihoods with little manual tuning, and truncation makes more efficient use of each simulation.

Compared to standard SNL, T-SNL offers several advantages. First, it is significantly more sample efficient as it is able to draw more likelihood information per sample. By leveraging the Markov property of the observation process, T-SNL transforms each simulation into multiple training examples, allowing it to achieve lower errors with fewer simulator calls. Moreover, the input size of the MAF model used by T-SNL does not depend on the sequence length $T$: the conditional $q_{\bphi}(\by_t|\by_{t-L:t-1}, \btheta)$ always sees a fixed-size input window. In contrast, SNL models the full likelihood and must process inputs of size proportional to $T$, which becomes increasingly difficult to scale. T-SNL not only avoids this issue, but also benefits from it in two distinct ways. Firstly, the lower dimensionality of the input size simplifies the learning problem and makes T-SNL more stable and robust during training, leading to fewer optimization failures and lower validation errors compared to SNL.  Secondly, independence from $T$ means that T-SNL can scale well to larger temporal sequences. Moreover, as $T$ increases the size of the training dataset grows proportionally, further improving performance. This can be seen in Fig. \ref{fig:val_loss} where the validation loss of SNL increases linearly with $T$ while for T-SNL it decreases. Finally, T-SNL naturally supports amortized inference: once the conditional likelihood model is trained, it can be reused for new observations without retraining. For example, if a new observation $\by_{T+1}$ becomes available, the learned conditional $q_{\bphi}(\by_{T+1}|\by_{T+1-L:T}, \btheta)$ can be used within an SMC or MCMC scheme without re-training. This is not the case for SNL where a new MAF model must be trained. 

Overall, our results highlight the importance of exploiting temporal structure in simulator-based inference for state-space models. By combining the flexibility of neural likelihood estimation with the Markovian structure imposed by the truncation, T-SNL makes better use of simulations and scales easily to longer sequences. The result is an inference method that is not only very sample-efficient, but also practical to apply in complex, real-world models. Moreover, the advantage of being amortized means that T-SNL can incorporate streaming information as it is obtained, updating the posterior estimate along the way. Therefore, it can be used as a reliable component of real-time inference algorithms, such as nested algorithms for joint state and parameter estimation \cite{perez2018probabilistic}.

\section{Conclusions}

In this work, we have introduced T-SNL, a novel, sample efficient variant of SNL. T-SNL takes advantage of the temporal structure of SSMs by using the forgetting property, which states that the influence of past observations is forgotten exponentially fast. The forgetting property, which is satisfied by many commonly used SSMs, allows T-SNL to replace the likelihood factors by their truncated version which only conditions on a fixed window of past observations. As a result T-SNL gains several distinct advantages. Firstly, it operates in a space of small and fixed dimensionality, making it more stable and robust during training. Secondly, in converts each simulation of the time-series into multiple training data points, thus it is trained on a much larger dataset than SNL. Moreover, the model size is independent from the length of the time-series and thus T-SNL scales well to large time-series. Finally, T-SNL provides better calibrated posterior samples and is amortized, allowing the trained model to be reused as new data arrives. Our experiments demonstrated that T-SNL outperforms other methods in both efficiency and robustness, particularly in complex and nonlinear models. Overall, our findings show that T-SNL is a flexible and effective tool for inference in SSMs, and suggest several promising directions for future work, including applications to real-time systems and more structured neural likelihood models.

{\small
\bibliography{bibliography}
}

\newpage

\setcounter{figure}{6}
\setcounter{algorithm}{1}

\begin{center}
{\large\textbf{Supplementary Material}}
\end{center}

\vspace{1em}

\section{SMC-ABC adaptive tolerance selection}
Here we derive our implementation of the adaptive tolerance selection method of \cite{abc_tolerance}. The implementation involves finding the ratio function and its maximization with respect to the parameters. Computationally this reduces to two optimization problems. We solve them using fixed point iterations, resulting in an efficient algorithm for tuning the tolerance adaptively.

\subsection{Derivation of tolerance selection subroutine}

\subsubsection*{Finding the ratio function}

The ratio is defined by:

\begin{align}
    r(\btheta) &= \frac{\widehat p_{\epsilon_{t}}(\btheta)}{\widehat p_{\epsilon_{t-1}}(\btheta)}, \\
    \widehat p_{\epsilon_{t}}(\btheta) &= \frac{1}{N}\sum_{n=1}^N \delta(\btheta-\btheta^{(n)}_{t}), \\
    \widehat p_{\epsilon_{t-1}}(\btheta) &= \frac{1}{N}\sum_{n=1}^N \delta(\btheta-\btheta^{(n)}_{t-1}).
\end{align}
We approximate the ratio by
\begin{equation}
        r_{\balpha}(\btheta) = \sum_{n=1}^N \alpha_n e^{-||\btheta-\btheta^{(n)}_{t-1}||^2 / 2\sigma^2}.
\end{equation}
In order for $r_{\balpha}$ to be a proper ratio, it must satisfy identically
\begin{align}
    1 &= \int r_{\balpha}(\btheta) \widehat p_{\epsilon_{t-1}}(\btheta) d\btheta \\
    &= \sum_{n=1}^N r_{\balpha}(\btheta_{t-1}^{(n)}) \\
    &= \sum_{n=1}^N \sum_{m=1}^N \alpha_{n} e^{-||\btheta^{(n)}_{t-1}-\btheta^{(m)}_{t-1}||^2 / 2\sigma^2} \\
    &= \sum_{n=1}^N \sum_{m=1}^N \alpha_{n} E_{nm}^{0} \\
    &= \sum_{n=1}^N \alpha_{n} \sum_{m=1}^N E_{nm}^{0} \\
    &= \balpha^T \be^0,
\end{align}
where
\begin{align}
    E_{nm}^{0} &= e^{-||\btheta^{(n)}_{t-1}-\btheta^{(m)}_{t-1}||^2 / 2\sigma^2} \\
    \be^0_n &= \sum_{m=1}^N E_{nm}^{0}.
\end{align}
We set $\balpha$ by maximizing the following function
\begin{align}
   \ell(\balpha) &= \sum_{n=1}^N \log r_{\balpha}(\btheta_{t}^{(n)}) - \lambda (1-\balpha^T\be^0)), \\
   &= \sum_n \log \sum_{m}\alpha_{m}E_{nm} - \lambda (1-\balpha^T\be^0)),
\end{align}
where $E_{nm} = e^{-||\btheta^{(n)}_{t}-\btheta^{(m)}_{t-1}||^2 / 2\sigma^2}$. We compute the gradient of this function
\begin{align}
    \frac{\partial \ell(\balpha)}{\partial\alpha_{k}}  &= \sum_n \frac{E_{nk}}{\sum_{m}\alpha_{m}E_{nm}} - \lambda \be^0_k,
\end{align}
If we multiply and divide the first term with $\alpha_k$ and set the gradient equal to zero we obtain
\begin{equation}
     \alpha_{k} = \frac{1}{\lambda \be^0_k} \sum_n \frac{\alpha_{k} E_{nk}}{\sum_{m}\alpha_{m}E_{nm}}
\end{equation}
Vectorizing we obtain
\begin{equation}
    \balpha = \balpha  \odot \frac{1}{N \be^0} \odot \bE^T\bbeta = \bff (\balpha)
\end{equation}
where $\odot$ denotes element-wise multiplication, $\bbeta = \frac{1}{\bE\balpha}$, $\bE$ is the matrix with elements $E_{nm}$ and $\lambda = N$ to satisfy the constraint.

We solve the optimization problem by repeating the iteration
\begin{equation}
    \balpha^{(t)} = \bff(\balpha^{(t-1)})
\end{equation}

\subsubsection*{Finding the supremum}
Our approximation of the ratio is $\widehat r_{\balpha^{\star}}(\btheta)$ and the corresponding value of $\widehat c_t$ is
\begin{equation}
    \widehat c_t = \sup_{\btheta}\widehat r_{\balpha^{\star}}(\btheta)
\end{equation}
To find it we follow a similar procedure to the previous subsection. Our objective to maximize is
\begin{equation}
    \widehat r_{\balpha^{\star}}(\btheta) = \sum_{n=1}^N \alpha^{\star}_n e^{-||\btheta-\btheta^{(n)}_{t-1}||^2 / 2\sigma^2}
\end{equation}
We take the gradient to obtain
\begin{align}
    \frac{\partial \widehat r_{\balpha^{\star}}(\btheta)}{\partial \btheta} &= -\frac{1}{2\sigma^2} \sum_n \alpha^{\star}_n \frac{\partial}{\partial \btheta} ||\btheta - \btheta_{t-1}^{(n)}||^2 e^{-||\btheta-\btheta^{(n)}_{t-1}||^2 / 2\sigma^2} \\
    &= -\frac{1}{\sigma^2} \sum_n \alpha^{\star}_n \frac{\partial}{\partial \btheta} (\btheta - \btheta_{t-1}^{(n)}) e^{-||\btheta-\btheta^{(n)}_{t-1}||^2 / 2\sigma^2}.
\end{align}
Setting equal to zero we obtain the fixed point equation
\begin{equation}
    \btheta = \frac{1}{\widehat r_{\balpha^{\star}}(\btheta)} \sum_n \alpha^{\star}_n e^{-||\btheta-\btheta^{(n)}_{t-1}||^2 / 2\sigma^2} \btheta_{t-1}^{(n)}
\end{equation}
which we abbreviate by $\btheta = \bg(\btheta)$.

\section{Model selection details}
The number of parameters $k(L)$ of a MAF model with $N_{m}$ MADE layers, each of which with $M$ hidden layers of $H$ hidden units, is equal to

\begin{equation}
    k(L) = N_{m} \times \Big( H \times (L+d_y+d_{\theta}+1) + (M-1)\times H^2 \Big).
\end{equation}
When $N_m=M=5$, $H=32$ and $d_\theta=dy=1$ we have $k(L) = 160L + 20960$.

\section{Relationship between $\mathcal{E}_{KDE}$ and $\mathcal{E}_{min}$}

Here we derive an approximation of $\mathcal{E}_{KDE}$ in terms of $\mathcal{E}_{min}$. Given samples $\{\widehat \btheta_k\}_{k=1}^K$ from the approximate posterior we use the Gaussian KDE
\begin{equation}
    p_{KDE}(\btheta) = \frac{1}{K}\sum_{k=1}^K \mathcal{N}(
    \btheta|\widehat \btheta_k, \sigma^2\bI).
\end{equation}
The log-pdf evaluated at the true parameters $\btheta_0$ can be approximated as follows
\begin{align}
    \log p_{KDE}(\btheta_0) &= \log \frac{1}{K}\sum_{k=1}^K \mathcal{N}(\btheta|\widehat \btheta_k, \sigma^2\bI) \\
    &= \log \frac{1}{K}\sum_{k=1}^K \frac{e^{-||\btheta_0-\btheta_k||^2/2\sigma^2}}{(2\pi\sigma^2)^{d_{\theta}/2}} \\
    &= \log \sum_{k=1}^K e^{-||\btheta_0-\btheta_k||^2/2\sigma^2} \nonumber \\
    & - \log K - \frac{d_{\theta}}{2}\log(2\pi\sigma^2) \label{LSE1} \\
    &= \text{LSE}(\{-\frac{1}{2\sigma^2}||\btheta_0-\btheta_k||^2\}_k) + C \nonumber
\end{align}
where $\text{LSE}$ is the log-sum-exp function and $C=- \log K - \frac{d_{\theta}}{2}\log(2\pi\sigma^2)$. To make the connection to $\mathcal{E}_{min}$ we use the following inequality which holds identically,
\begin{equation} \label{LSE2}
    \max\{e_k\}_{k=1}^K \leq \text{LSE}(\{e_k\}_{k=1}^K) \leq  \max\{e_k\}_{k=1}^K + \log K.
\end{equation}
Combining Eqs. \eqref{LSE1} and \eqref{LSE2} we obtain for the error $\mathcal{E}_{KDE}$:
\begin{equation}
    \resizebox{\columnwidth}{!}{$\displaystyle\frac{1}{2\sigma^2}\mathcal{E}_{min}^2 + \frac{d_{\theta}}{2}\log(2\pi\sigma^2) \leq \mathcal{E}_{KDE} \leq \frac{1}{2\sigma^2}\mathcal{E}_{min}^2 + \log K + \frac{d_{\theta}}{2}\log(2\pi\sigma^2)$}
\end{equation}

\end{document}